\documentclass[%
reprint,
 superscriptaddress,
 nofootinbib,
 amsmath,amssymb,
 aps,
]{revtex4-2}

\usepackage{graphicx}
\usepackage{dcolumn}
\usepackage{bm}
\usepackage{hyperref}
\usepackage[mathlines]{lineno}

\usepackage{bbold}

\newcommand{\beq}{\begin{equation}}
\newcommand{\eeq}{\end{equation}}
\newcommand{\bea}{\begin{eqnarray}}
\newcommand{\eea}{\end{eqnarray}}

\newcommand{\Tr}{\mathop{\rm Tr}}

\newcommand{\Slash}[1]{{\ooalign{\hfil#1\hfil\crcr\raise.167ex\hbox{/}}}}

\begin{document}


\title{Truth, beauty, and goodness in grand unification: a machine learning approach}

\author{Shinsuke Kawai}
	\email{kawai@skku.edu}
	\affiliation{%
	Department of Physics, Sungkyunkwan University, 
	Suwon, 16419 Republic of Korea}
 
\author{Nobuchika Okada}
 \email{okadan@ua.edu}
 \affiliation{%
Department of Physics and Astronomy, University of Alabama, Tuscaloosa, Alabama, AL35487 USA}


\begin{abstract}
We investigate the flavour sector of the supersymmetric $SU(5)$ Grand Unified Theory (GUT) model using machine learning techniques. 
The minimal $SU(5)$ model is known to predict fermion masses that disagree with observed values in nature. 
There are two well-known approaches to address this issue: one involves introducing a 45-representation Higgs field, while the other employs a higher-dimensional operator involving the 24-representation GUT Higgs field. 
We compare these two approaches by numerically optimising a loss function, defined as the ratio of determinants of mass matrices. 
Our findings indicate that the 24-Higgs approach achieves the observed fermion masses with smaller modifications to the original minimal $SU(5)$ model.
\end{abstract}

\maketitle


\section{\label{sec:intro}Introduction}

Unification of forces at high energies is a key concept in particle physics. 
This is supported by the success of the Standard Model, which unifies the electromagnetic and weak interactions under the gauge principle. 
The renormalisation group flows of the three gauge couplings in the minimal supersymmetric Standard Model (MSSM) further suggest that grand unification of the strong and electroweak interactions occurs around $10^{16}$ GeV. 
Whether this unification happens in nature, and if so, how it is described by a grand unified theory (GUT), are primary concerns of physics beyond the Standard Model. 
Grand unification has also been the cradle of many of today's dominant paradigms in high-energy physics, including cosmic inflation and baryogenesis. 
The study of GUT itself, particularly the fixing of parameters against experimental constraints, remains an important direction of research.

The simplest realisation of GUT, the minimal $SU(5)$ model \cite{Georgi:1974sy}, is in fact excluded by experiments.
The model organises the leptons and quarks of the Standard Model into the $\bm{10}$ and $\overline{\bm{5}}$ representations of $SU(5)$, which leads to relations among the masses of the down-type quarks and the charged leptons at the GUT energy scale,
\begin{align}\label{eqn:GUTmassRel}
	m_d=m_e,\quad
	m_s=m_\mu,\quad
	m_b=m_\tau.
\end{align} 
Renormalisation group analysis clearly shows that this is inconsistent with the fermion masses that we observe.
One way of circumventing this difficulty and accommodating the observed fermion masses in the $SU(5)$ model is to enlarge the Higgs sector and consider coupling of the fermions to a $\overline{\bm{45}}$ representation field of $SU(5)$ \cite{Georgi:1979df}.
Another attitude is to suppose that the mass relation at the GUT scale is broken as the mass spectrum can be susceptible to physics at the fundamental scale, e.g. the Planck scale.
For example, nonrenormalisable coupling of the fermions to the $\bm{24}$ representation Higgs (which is necessary to break the GUT symmetry in the $SU(5)$ model) successfully gives the observed fermion masses \cite{Ellis:1979fg}.
In both approaches, new (effective) Yukawa coupling in the form of a $3\times 3$ complex matrix is introduced, and the observed fermion masses may be obtained by adjusting the new parametric degrees of freedom, which comes at the expense of certain loss of predictability.

We frequently encounter similar situations in physics, where an original theoretical model, characterised by simplicity, predictability and {\em beauty} in some sense, is too restrictive to explain experimental data, or the {\em truth}.
To reconcile with reality, new elements are added to the model, allowing it to fit the experimental data.
However, the parameter space often becomes so large that a comprehensive parameter search is impossible.  
In practice, researchers are not interested in all sets of parameters that fit the data, but only in those sets of {\em good} parameters that approximately conform to the principles of the original {\em beautiful} model.
Recent studies \cite{Halverson:2019tkf,Nishimura:2020nre,Harvey:2021oue,Cole:2021nnt,deSouza:2022uhk,Dersy:2023job,Romao:2024gjx,Nishimura:2024apb}, particularly \cite{Matchev:2024ash}, show that techniques of machine learning can be usefully applied to handle this type of problem.
The machine learning approach differs somewhat from the traditional theoretical approach; instead of comprehensive investigation of parameters, a vast parameter space is explored through sampling and optimisation.

In this paper, we revisit the fermion mass problem of the $SU(5)$ GUT model using machine learning techniques. 
For the sake of concreteness we shall discuss the supersymmetric $SU(5)$ GUT model as it has various merits over the nonsupersymmetric counterpart, including technical naturalness and crisp gauge coupling unification.
In next section we review the supersymmetric $SU(5)$ model and in Sec.~\ref{sec:gutmm} the fermion mass matrix at the GUT scale is derived.
Our method of analysis based on machine learning techniques is explained in Sec.~\ref{sec:ML} and the numerical results are presented in Sec.~\ref{sec:results}. 
We conclude with brief comments in Sec.~\ref{sec:final}.

\section{\label{sec:su5gut} supersymmetric $SU(5)$ model}

We start with the minimal $SU(5)$ model \cite{Georgi:1974sy}.
It consists of one $\bm{24}$ representation vector supermultiplet $V_{\bm{24}}$, and one $\bm{24}$, one $\bm{5}$, one $\bm{\overline 5}$ and $N_F=3$ copies of $\bm{10}$ and $\bm{\overline 5}$ representation chiral supermultiplets 
	that we denote by $H_{\bm{24}}$, $H_{\bm{5}}$, $H_{\bm{\overline 5}}$, $F^i_{\bm{10}}$ and $F^i_{\bm{\overline 5}}$.
The family indices $i, j, ... \in \{1, ..., N_F\}$ will be suppressed unless necessary.
The superpotential of this model is
\begin{align}
	W_{\rm min}=&\frac M2 \Tr\left( H_{\bm{24}} \right)^2
	+\frac{\Lambda}{3} \Tr\left(H_{\bm{24}}\right)^3 
	+H_{\bm{\overline 5}} \left(\mu+\lambda H_{\bm{24}}\right)H_{\bm{5}}\nonumber\\
	&-y^{5d}_{ij}[F^i_{\bm{\overline 5}}]^m[F^j_{\bm{10}}]_{mn}[H_{\bm{5}}]^n\nonumber\\
	&-y^{5u}_{ij}\epsilon^{mnpqr}[F^i_{\bm{10}}]_{mn}[F^j_{\bm{10}}]_{pq}[H_{\bm{\overline 5}}]_r,
\end{align}
where $\epsilon$ is the antisymmetric tensor, $y^{5u}_{ij}$ and $y^{5d}_{ij}$ are the Yukawa couplings, 
$m,n,...\in\{1, 2, 3, 4, 5\}$ are the $SU(5)$ indices, and $M$, $\Lambda$, $\mu$ and $\lambda$ are parameters of the model that may be adjusted to have successful symmetry breaking $SU(5)\to SU(3)_c\times SU(2)_L\times U(1)_Y\to SU(3)_c\times U(1)_{EM}$.
The quarks and leptons of the Standard Model are embedded in $F_{\bm{\overline 5}}$ and $F_{\bm{10}}$ as
\begin{align}\label{eqn:F5}
	[F_{\bm{\overline 5}}]^m=\begin{bmatrix}
		d_1^c,\; d_2^c,\; d_3^c,\; e,\; -\nu_e
	\end{bmatrix},
\end{align}
\begin{align}\label{eqn:F10}
	[F_{\bm{10}}]_{mn}=\frac{1}{\sqrt 2}\begin{bmatrix}
		0 & u_3^c & -u_2^c & -u_1 & -d_1\\
		-u_3^c & 0 & u_1^c & -u_2 & -d_2\\
		u_2^c & -u_1^c & 0 & -u_3 & -d_3\\
		u_1 & u_2 & u_3 & 0 & -e^c \\
		d_1 & d_2 & d_3 & e^c & 0
	\end{bmatrix}
\end{align}
where the lower indices 1, 2, 3 of the quarks $u$, $d$ represent the colour.

\subsection{\label{sec:minsu5}GUT mass relation of the minimal $SU(5)$ model}

Below the electroweak symmetry breaking scale, the $\bm{5}$ and $\bm{\overline 5}$ Higgs acquire vacuum expectation values
\begin{align}\label{eqn:H5vevs}
	\langle H_{\bm{5}}\rangle=[0, 0, 0, 0, \frac{v_u}{\sqrt 2}], \qquad
	\langle H_{\bm{\overline 5}}\rangle=[0, 0, 0, 0, \frac{v_d}{\sqrt 2}].
\end{align}
The Yukawa part of the Lagrangian then becomes
\begin{align}
	y^{5d}_{ij}\frac{v_d}{\sqrt 2}[\overline{\Psi^i_{\bm{\overline 5}}}]^m[\Psi^j_{\bm{10}}]_{m5}
	&+y^{5u}_{ij}\frac{v_u}{\sqrt 2}\epsilon^{mnpq5}[\overline{\Psi^i_{\bm{10}}}]_{mn}[\Psi^j_{\bm{10}}]_{pq}\nonumber\\
	&\qquad\qquad\qquad\qquad\qquad+{\rm h.c.},
\end{align}
where $\Psi^i_{\bm{\overline 5}}$, $\Psi^i_{\bm{10}}$ are the fermionic part of $F^i_{\bm{\overline 5}}$ and $F^i_{\bm{10}}$.
The mass matrices are found to be
\begin{align}\label{eqn:M5rel}
	M_u =4\frac{v_u}{\sqrt 2} y^{5u}_{ij},\;\;
	M_d = \frac{v_d}{2} y^{5d}_{ij},\;\;
	M_e = \frac{v_d}{2} y^{5d}_{ji} = M_d^T.
\end{align}
The minimal $SU(5)$ model thus predicts the relation for the fermion masses \eqref{eqn:GUTmassRel} at the GUT scale.
This is a strong constraint on the renormalisation group flow of the Yukawa couplings, and since it is not satisfied, one needs to go beyond the minimal $SU(5)$ model.

\subsection{\label{sec:H45}Extension with 45-Higgs}

One approach to amend this fermion mass relation \cite{Georgi:1979df} is by extending the Higgs sector with $\bm{\overline{45}}$ representation field $H_{\bm{\overline{45}}}$.
In the supersymmetric theory that we consider, its partner $H_{\bm{45}}$ belonging to the $\bm{45}$ representation is also introduced to form a vector-like pair. 
These are represented by 
$[H_{\overline{\bm{45}}}]^{np}_m$ and
$[H_{\bm{45}}]^{m}_{np}$ such that
$[H_{\overline{\bm{45}}}]^{np}_m=-[H_{\overline{\bm{45}}}]^{pn}_m$, 
$[H_{\overline{\bm{45}}}]^{mp}_m=0$ and 
$[H_{\bm{45}}]^{m}_{np}=-[H_{\bm{45}}]^{m}_{pn}$, $[H_{\bm{45}}]^{m}_{mp}=0$.
Introducing a new term of the superpotential\footnote{A term of the form
$\epsilon^{mnprs}[F^i_{\bm{10}}]_{mn}[F^j_{\bm{10}}]_{pq}[H_{\bm{45}}]^{q}_{rs}$
may also be included, but this will not affect the mass relation we discuss.
}
\begin{align}
	W_{45}=&\;y^{45d}_{ij}[F^i_{\bm{\overline 5}}]^m[F^j_{\bm{10}}]_{np}[H_{\overline{\bm{45}}}]^{np}_m
\end{align}
and assumimg $SU(3)_c\times U(1)_Y$-invariant vacuum expectation value
\begin{align}
	\langle[H_{\overline{\bm{45}}}]^{n5}_m\rangle = v_{45}\;{\rm diag}(1, 1, 1, -3, 0),
\end{align}
the mass matrix may be computed.
Combined with the contributions from the minimal $SU(5)$ part, one finds
\begin{align}\label{eqn:M45rel}
	M_u =& 2\sqrt 2 v_u y^{5u}_{ij},\nonumber\\
	M_d =& \frac{v_d}{2} y^{5d}_{ij}+ \frac{v_{45}}{\sqrt 2}\, y^{45d}_{ij},\nonumber\\
	M_e =& \frac{v_d}{2} y^{5d}_{ji}-3\frac{v_{45}}{\sqrt 2}\, y^{45d}_{ji}.
\end{align}
This breaks the relation \eqref{eqn:GUTmassRel}.
The Yukawa matrices $y^{5d}_{ij}$ and $y^{45d}_{ij}$ may be adjusted to accommodate the observed fermion mass spectrum.

\subsection{\label{sec:H24}Extension by nonrenormalisable coupling with 24-Higgs}%

An alternative approach is to consider contributions from higher dimensional operators \cite{Ellis:1979fg}.
$SU(5)$ gauge singlet $F_{\bm{\overline 5}}H_{\bm{24}}F_{\bm{10}}H_{\bm{\overline 5}}\subset W$ gives Lagrangian of the form
\begin{align}
	\frac{y^{24d}_{ij}}{m_{\rm P}}[\overline{\Psi^i_{\bm{\overline 5}}}]^m\langle [H_{\bm{24}}]_m{}^n\rangle [\Psi^j_{\bm{10}}]_{np}\langle [H_{\bm{\overline 5}}]^p\rangle+{\rm h.c.}, 
\end{align}
where $y^{24d}_{ij}$ is a new Yukawa-like coupling and $m_{\rm P}$ is the Planck mass.
Upon GUT symmetry breaking, $H_{\bm{24}}$ obtains vacuum expectation value
$H_{\bm{24}}=(M/\Lambda)\,{\rm diag}(2,2,2,-3,-3)$.
The expectation value of $H_{\bm{\overline 5}}$ is as given by \eqref{eqn:H5vevs}.
The fermion mass matrices then become, including the minimal $SU(5)$ contributions,
\begin{align}\label{eqn:M24rel}
	M_u =& 2\sqrt 2 v_u y^{5u}_{ij},\nonumber\\
	M_d =& \frac{v_d}{2} y^{5d}_{ij}+ \frac{M}{m_{\rm P}\Lambda}\frac{v_d}{\sqrt 2}\, y^{24d}_{ij},\nonumber\\
	M_e =& \frac{v_d}{2} y^{5d}_{ji}- \frac{3}{2}\frac{M}{m_{\rm P}\Lambda}\frac{v_d}{\sqrt 2}\, y^{24d}_{ji}.
\end{align}
Also in this case, it is known that the observed fermion mass spectrum is obtained by adjusting $y^{5d}_{ij}$ and $y^{24d}_{ij}$.

\section{\label{sec:gutmm}Fermion mass matrix at the GUT scale}

We recall that a Yukawa matrix is diagonalised by two unitary matrices that we denote by $V$ and $U$.
Explicitly, 
\begin{align}
	\frac{v_u}{\sqrt 2}V_u^\dag y^u U_u = {\rm diag}(m_u, m_c, m_t)\equiv D_u,\nonumber\\
	\frac{v_d}{\sqrt 2}V_d^\dag y^d U_d = {\rm diag}(m_d, m_s, m_b)\equiv D_d,\nonumber\\
	\frac{v_d}{\sqrt 2}V_e^\dag y^e U_e = {\rm diag}(m_e, m_\mu, m_\tau)\equiv D_e.
\end{align}
The neutrino Dirac Yukawa matrix may also be diagonalised similarly, but we will not discuss it below since the values of the neutrino Yukawa coupling are not constrained by experiments. 
This disregard of the neutrino Yukawa coupling is justified if the seesaw scale is sufficiently high and not very far from the GUT scale.
The Cabibbo–Kobayashi–Maskawa (CKM) matrix is
\begin{align}\label{eqn:CKMdef}
	V_{\rm CKM}=V_u^\dag V_d.
\end{align}
The mass matrices that appeared in \eqref{eqn:M5rel}, \eqref{eqn:M45rel} and \eqref{eqn:M24rel} are 
\begin{align}\label{eqn:Mdiag}
	M_u\equiv &\frac{v_u}{\sqrt 2}y^u = V_uD_uU_u^\dag,\nonumber\\
	M_d\equiv &\frac{v_d}{\sqrt 2}y^d = V_dD_dU_d^\dag,\nonumber\\
	M_e\equiv &\frac{v_d}{\sqrt 2}y^e = V_eD_eU_e^\dag,
\end{align}
evaluated at the GUT scale.
Diagonalisation of the Yukawa matrices is carried out separately at the electroweak scale and at the GUT scale.
The diagonalising matrices $V$ and $U$ at the GUT scale take different values from those at the electroweak scale, as the Yukawa couplings evolve under the renormalisation group flow.
In particular, the CKM matrix \eqref{eqn:CKMdef} at the GUT scale is different from the one at the low energy.

We wish to explore the parameter space of the flavour sector of the $SU(5)$ GUT, for the two models of extension beyond the minimal $SU(5)$, under the condition that they both predict the observed fermion masses at low energy. 
For that, we need to solve the renormalisation group equations using the fermion masses at low energy up to the GUT scale, as described below.
We assume the scenario of low scale supersymmetry breaking for the sake of concreteness.

At the one loop level, the renormalisation group equations for the gauge couplings $g_i$, $i=1,2,3$ of the $U(1)_Y$, $SU(2)_L$, $SU(3)_c$ groups are 
\begin{align}
	16\pi^2\frac{dg_i}{d\ln\mu}=-b_i g_i^3,
\end{align}
where $\mu$ is the renormalisation scale and $b_i\equiv (b_1, b_2, b_3)=(-33/5,\, -1,\, 3)$ for the MSSM.
We use boundary conditions $\alpha_i\equiv g_i^2/4\pi=(0.0168,\; 0.0335,\; 0.118)$ at $\mu=M_Z=91.2$ GeV\footnote{
As long as the supersymmetry breaking scale is $M_S\simeq 10$ TeV for which the Super-Kamiokande bounds \cite{ParticleDataGroup:2024cfk} on the proton decay through the 5-dimensional operator are satisfied, the GUT scale output does not depend much on whether the renormalisation group flow between $\mu=M_Z$ and $\mu=M_S$ is that of the Standard Model or of the MSSM.
}.
Solving them toward high energies, the GUT scale is found as $M_U\simeq 3\times 10^{16}$ GeV.

The renormalisation group equations for the Yukawa couplings $y^f$, $f=\{u, d, e\}$, may be arranged in the form 
\begin{align}
	16\pi^2\frac{dS_f}{d\ln\mu}=\beta_fS_f+S_f\beta_f,
\end{align}
where $S_f\equiv (y^f)^\dag y^f$ and \cite{Castano:1993ri}
\begin{align}
	\beta_u=&3S_u+S_d+\left\{\Tr(3S_u)-\frac{13}{15}g_1^2-3g_2^2-\frac{16}{3}g_3^2\right\}{\mathbb{1}},\nonumber\\
	\beta_d=&3S_d+S_u+\left\{\Tr(3S_d+S_e)-\frac{7}{15}g_1^2-3g_2^2-\frac{16}{3}g_3^2\right\}{\mathbb{1}},\nonumber\\
	\beta_e=&3S_e+\left\{\Tr(3S_d+S_e)-\frac{9}{5}g_1^2-3g_2^2\right\}{\mathbb{1}}.
\end{align}
The input values at low energy $\mu=M_Z$ are given by
\begin{align}
	S_u =& {\rm diag}\left(\frac{m_u^2}{v^2}, \frac{m_c^2}{v^2}, \frac{m_t^2}{v^2}\right)\left(1+\frac{1}{\tan^2\beta}\right),\nonumber\\
	S_u =& V_{\rm CKM}{\rm diag}\left(\frac{m_d^2}{v^2}, \frac{m_s^2}{v^2}, \frac{m_b^2}{v^2}\right)V_{\rm CKM}^\dag (1+\tan^2\beta),\nonumber\\
	S_u =& {\rm diag}\left(\frac{m_e^2}{v^2}, \frac{m_\mu^2}{v^2}, \frac{m_\tau^2}{v^2}\right)\left(1+\frac{1}{\tan\beta}\right),
\end{align}
in the basis in which $S_u$ and $S_e$ are diagonal.
We use $\tan\beta\equiv v_d/v_u = 10$ and the mean values of the fermion masses \cite{Ohlsson:2018qpt}
\begin{widetext}
\begin{align}
  	m_u(M_Z) &= 0.00127,  & m_c(M_Z)   &= 0.634,  & m_t(M_Z)    &= 171, \nonumber\\
	m_d(M_Z) &= 0.00271,  & m_s(M_Z)   &= 0.0553, & m_b(M_Z)    &= 2.86, \nonumber\\
	m_e(M_Z) &= 0.000487, & m_\mu(M_Z) &= 0.103,  & m_\tau(M_Z) &= 1.75, 
\end{align}
in GeV, and the CKM parameters 
\begin{align}
	s_{12}(M_Z)&=0.225,   & s_{23}(M_Z)&=0.0411, &
	s_{13}(M_Z)&=0.00357, & \delta_{13}(M_Z) &= 1.24
\end{align}
for the standard parametrisation of the CKM matrix (i.e. the form of \eqref{eqn:CKMlike} below).
Solving the renormalisation group equations, the fermion mass parameters at the GUT scale $\mu=M_U$ are
\begin{align}\label{eqn:GUTmass}
  	m_u(M_U) &= 0.000502, & m_c(M_U)   &= 0.251,  & m_t(M_U)    &= 97.7, \nonumber\\
	m_d(M_U) &= 0.000769, & m_s(M_U)   &= 0.0157, & m_b(M_U)    &= 0.922, \nonumber\\
	m_e(M_U) &= 0.000324, & m_\mu(M_U) &= 0.0685, & m_\tau(M_U) &= 1.17, 
\end{align}
in GeV.
The CKM matrix elements at the GUT scale are found to be
\begin{align}\label{eqn:GUTCKM}
V_{\rm CKM}(M_U)=
	\begin{pmatrix}
		0.973 + 0.0534 i &
		0.211 - 0.0769 i &
		2.19\times 10^{-6} - 0.00315 i\\
		-0.208 - 0.0845 i &
		0.974 - 0.0184 i & 
		0.0363 - 1.91\times 10^{-7} i\\
		0.00774 - 1.35\times 10^{-10} i&
		-0.0356 - 2.94\times 10^{-11} i& 
		0.999 - 3.33\times 10^{-16} i
	\end{pmatrix},
\end{align}
\end{widetext}
in the standard parametrisation.

\section{\label{sec:ML}Numerical exploration of the flavour sector of the $SU(5)$ model}

We reviewed in Sec.~\ref{sec:su5gut} the two well known approaches of amending the GUT mass relation.
One is by extending the model with the new Higgs field belonging to the $\bm{\overline{45}}$-representation of $SU(5)$, which we shall call the 45-Higgs model.
The other approach considers a higher dimensional operator involving the $\bm{24}$-representation Higgs.
We shall call this the 24-Higgs model.
These two approaches give different predictions \eqref{eqn:M45rel} and \eqref{eqn:M24rel} of the mass matrices.
Denoting 
\begin{align*}
	M_5 \equiv  \frac{v_d}{2} y^{5d}_{ij},\quad
	M_{45} \equiv \frac{v_{45}}{\sqrt 2}\, y^{45d}_{ij},\quad
	M_{24} \equiv \frac{M}{2m_{\rm P}\Lambda}\frac{v_d}{\sqrt 2}\, y^{24d}_{ij},
\end{align*}
the mass relations are written 
\begin{align}\label{eqn:M45rel2}
	M_d &= M_5+M_{45},\nonumber\\
	M_e &= M_5^T-3M_{45}^T
\end{align}
for the 45-Higgs model, and 
\begin{align}\label{eqn:M24rel2}
	M_d &= M_5+2M_{24},\nonumber\\
	M_e &= M_5^T-3M_{24}^T
\end{align}
for the 24-Higgs model.

Both of these systems have sufficient parametric degrees of freedom to reproduce the experimentally observed values of the quark and lepton mass eigenvalues, as well as of the CKM matrix elements. 
The remaining space of free parameters is still vast, making an exhaustive parameter search impossible.
However, we are typically not interested in the entire parameter space.
Our interest lies in parameters that correspond to the theoretical model remaining close to the simple but experimentally excluded original minimal $SU(5)$ model. 
The closeness to the minimal $SU(5)$ model is certainly a desirable feature for the 24-Higgs model, as it makes use of the higher-dimensional operator.
The relation \eqref{eqn:M45rel2} is written as
\begin{align}
	M_5 &= \frac 14 (3M_d+M_e^T),\nonumber\\
	M_{45} &= \frac 14 (M_d-M_e^T),
\end{align}
and the closeness to the minimal $SU(5)$ means smallness of the mass matrix $M_{45}$ relative to $M_5$ in some way.
As a simple criterion of this condition satisfying invariance under unitary transformations, we define the ratio of the determinants
\begin{align}\label{eqn:L45}
	L_{45}\equiv\frac{|\det M_{45}|}{|\det M_5|}
	=\left|\frac{\det (M_d-M_e^T)}{\det (3M_d+M_e^T)}\right|
\end{align}
and consider minimising it.
By the same token, in the case of the 24-Higgs model we consider minimisation of 
\begin{align}\label{eqn:L24}
	L_{24}\equiv\frac{|\det M_{24}|}{|\det M_5|}
	=\left|\frac{\det (M_d-M_e^T)}{\det (3M_d+2M_e^T)}\right|.
\end{align}
%

\subsection{\label{sec:params}Parametrisation}%

\begin{figure}
\includegraphics[width=95mm]{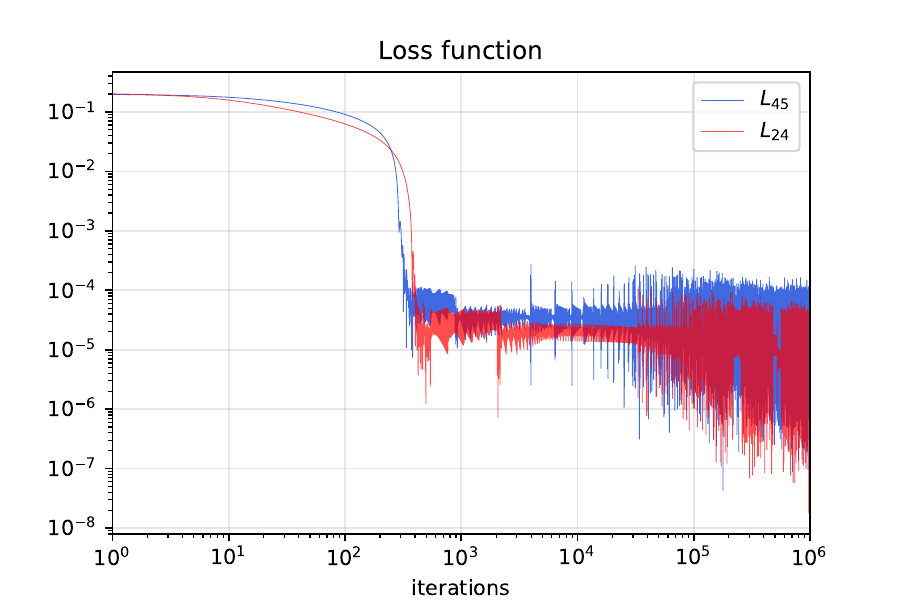}%
\caption
{\label{fig:LossEvo}
Behaviour of the loss function upon optimisation.
A sample of the loss function for the 45-Higgs model (24-Higgs model) is shown in blue (red).
Same initial parameters are used, and the optimisation is made up to $N_{\rm iter}=10^6$ iteration steps.
}
\end{figure}

Numerical optimisation of a {\rm loss function} is an essential technique in machine learning.
Considering $L_{45}$ and $L_{24}$ as loss functions, we may use this methodology for our minimisation problem. 
To proceed, we need to parametrise $M_e$ and $M_d$ appearing in \eqref{eqn:L45} and \eqref{eqn:L24}, ensuring that the constraints from the low energy fermion masses are satisfied.

In \eqref{eqn:Mdiag}, we may choose a basis in which $M_u$ and $M_e$ are diagonal:
\begin{align}
	M_u &= D_u = {\rm diag}(m_u, m_c, m_t),\nonumber\\
	M_e &= D_e = {\rm diag}(m_e, m_\mu, m_\tau).
\end{align}
Then $M_d$ cannot be made diagonal, and using $V_d=V_uV_{\rm CKM}$ from \eqref{eqn:CKMdef} we may write
\begin{align}\label{eqn:Md}
	M_d = V_uV_{\rm CKM}D_dU_d^\dag.
\end{align}
Here, $V_u$ is diagonal, but $U_d^\dag$ is an arbitrary unitary matrix.
We use parameterisation of a $3\times 3$ unitary matrix by 9 real variables
\begin{align}
	&U(\phi_0,\phi_1,\phi_2,\theta_1,\theta_2,\delta, \theta_3, \chi_1, \chi_2)\nonumber\\
	&=e^{i\phi_0} e^{i(\phi_1\lambda_3+\phi_2\lambda_8)}
	R(\theta_1,\theta_2,\delta,\theta_3) e^{i(\chi_1\lambda_3+\chi_2\lambda_8)},
\end{align}
where 
\begin{align}\label{eqn:CKMlike}
	&R(\theta_1,\theta_2,\delta,\theta_3)\\
	&=\begin{pmatrix}1&0&0\\ 0&c_{1}&s_{1}\\ 0&-s_{1}&c_{1}\end{pmatrix}
	\begin{pmatrix}c_{2}&0&s_{2}e^{-i\delta}\\ 0&1&0\\ -s_{2}e^{i\delta}&0&c_{2}\end{pmatrix}
	\begin{pmatrix}c_{3}&s_{3}&0\\ -s_{3}&c_{3}&0\\ 0&0&1\end{pmatrix}\nonumber
\end{align}
is a CKM-like matrix, with $s_i\equiv\sin\theta_i$, $c_i\equiv\cos\theta_i$ and 
$\lambda_3={\rm diag}(1, -1, 0)$, $\lambda_8={\rm diag}(1, 1, -2)/\sqrt 3$ (the Cartan part of the Gell-Mann matrices).

The mass matrices $M_e$ and $M_d$ appearing in the loss functions $L_{45}$ and $L_{24}$ are evaluated at the GUT scale.
Namely, we use
\begin{align}
	M_e =M_e(M_U) = (m_e(M_U), m_\mu(M_U), m_\tau(M_U)),
\end{align}
with the mass parameters at the GUT scale given by \eqref{eqn:GUTmass}.
Using \eqref{eqn:CKMlike} for $U_d^\dag$ and rearranging the communing diagonal elements in \eqref{eqn:Md}, $M_d$ is parametrised as
\begin{align}\label{eqn:MdParam}
	M_d &=M_d(M_U) \nonumber\\
	&= e^{ix_0}\,e^{i(x_1\lambda_3+x_2\lambda_8)}\,V_{\rm CKM}(M_U)\,D_d(M_U)\nonumber\\
	&\;\times e^{i(x_3\lambda_3+x_4\lambda_8)} R(x_5, x_6, x_7, x_8)\, e^{i(x_9\lambda_3+x_{10}\lambda_8)}.
\end{align}
Here, the CKM matrix at the GUT scale is \eqref{eqn:GUTCKM} and
\begin{align}
	D_d(M_U) = {\rm diag}(m_d(M_U), m_s(M_U), m_b(M_U)),
\end{align}
with the fermion masses at the GUT scale given in \eqref{eqn:GUTmass}.
The 11 parameters $x_0, \cdots, x_{10}$ are all real and are assumed to take initial values $0\leq x_i<2\pi$.

\begin{figure}
\includegraphics[width=85mm]{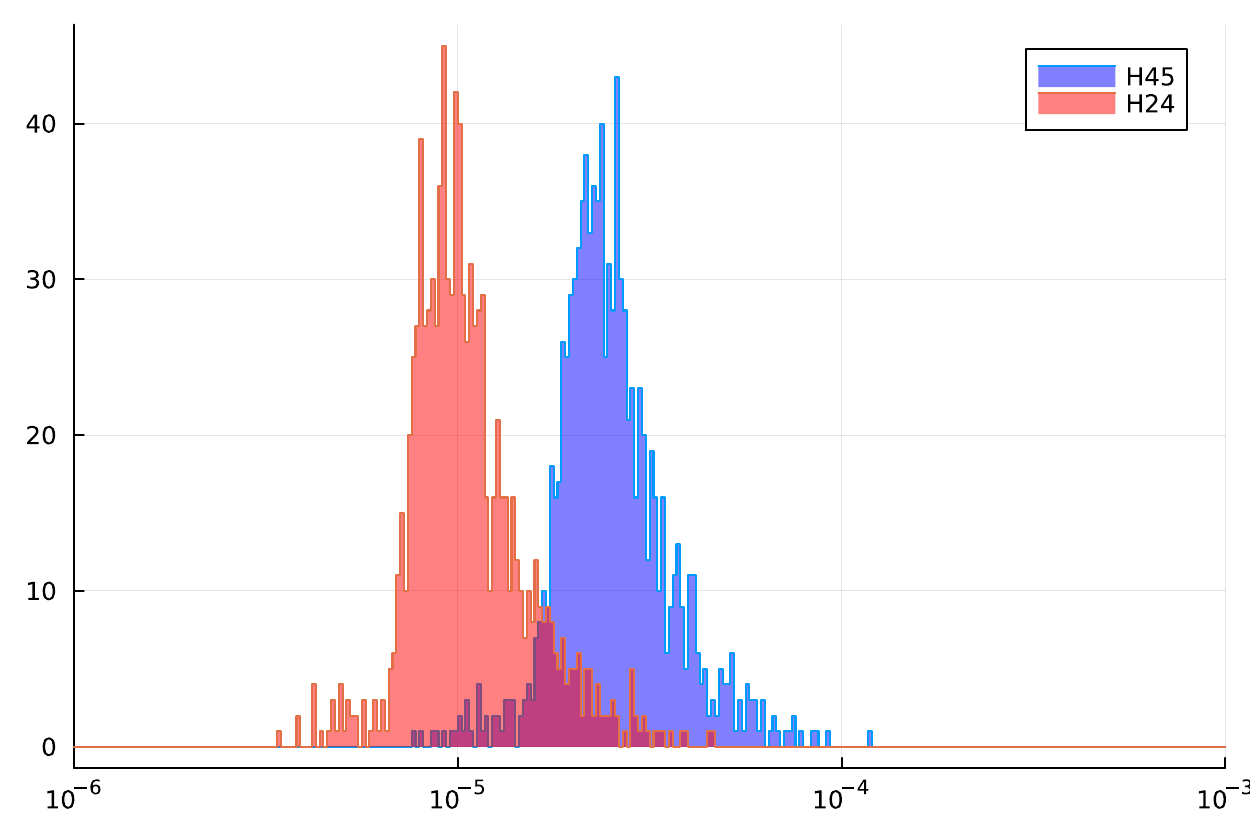}%
\caption
{\label{fig:histogram}
Distribution of the loss function values after optimisation, for the 45-Higgs model (H45, blue) and the 24-Higgs model (H24, red).
Optimisation is made for $N_{\rm iter}=10^6$ iterations, and $N_{\rm samp} = 1000$ samples are collected for each model.
The distribution of the loss function for the 24-Higgs model is seen to be peaked at a smaller value than that of the 45-Higgs model. 
}
\end{figure}

\subsection{\label{sec:SampOptim}Sampling and optimisation}%

\begin{figure*}
\includegraphics[width=85mm]{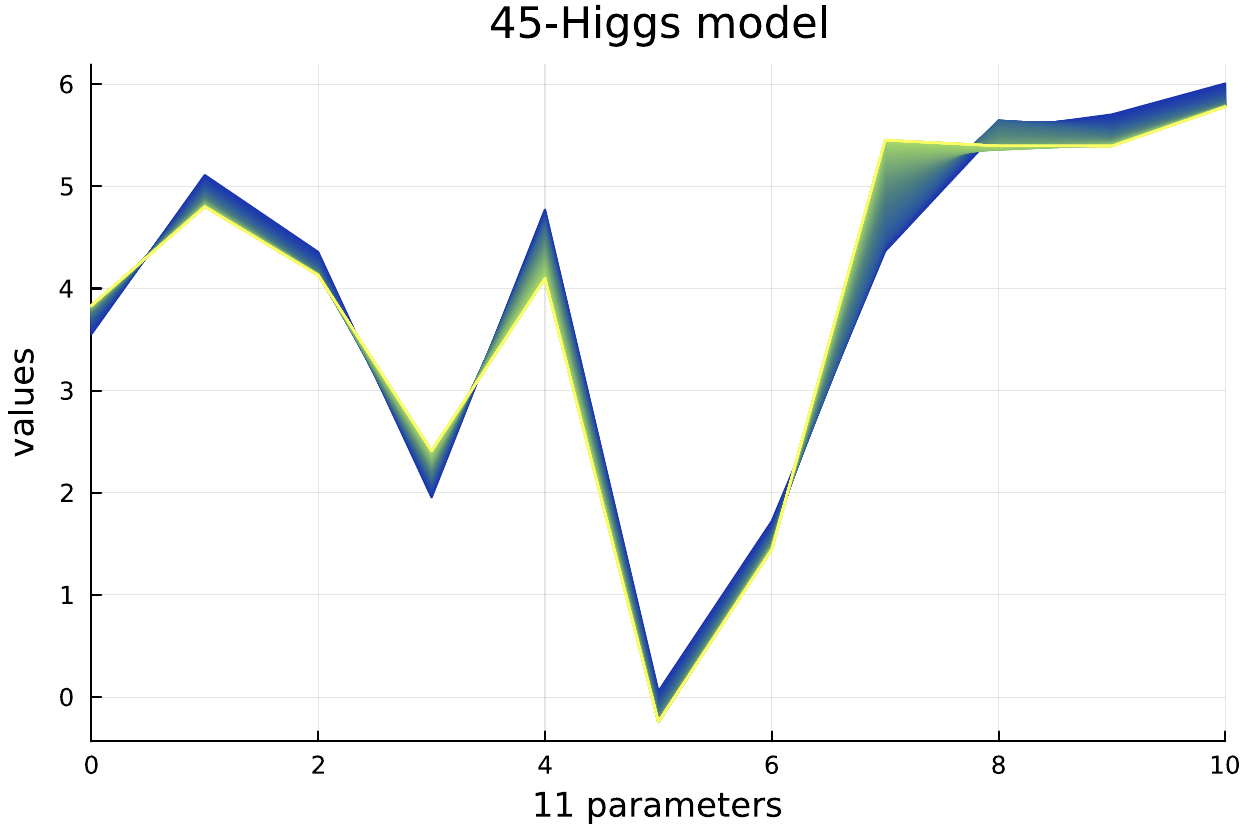}%
\includegraphics[width=85mm]{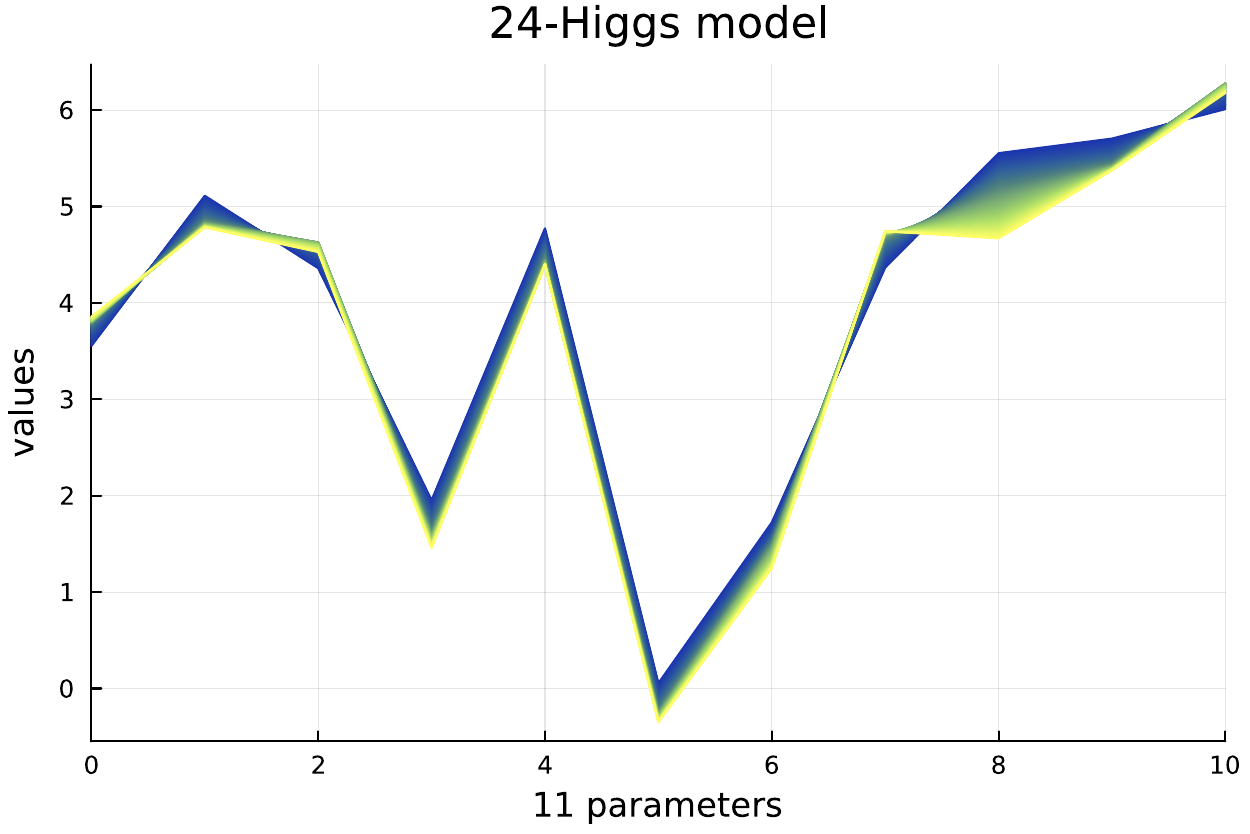}%
\caption
{\label{fig:ParamEvo}
Example of the evolution of the eleven parameters $x_i$, $i=0,2,...,10$ in the optimisation process. 
The darkest blue lines represent the initial random values of the parameters ($N_{\rm iter}=0$), and the brightest yellow lines represent the parameter configuration at $N_{\rm iter}=1,000$, after which the change of the parameter values is found to be small.
The lines in-between show parameter configurations at an interval of 10 iteration steps.
The left and the right panels show the results for the 45-Higgs and 24-Higgs model, starting with a same initial parameter configuration.
The parameters of the two models are seen to evolve differently, toward different optimised configurations. 
}
\end{figure*}

The problem we wish to tackle is to find a set of parameters $x_0,\cdots, x_{10}$ that minimises the loss function \eqref{eqn:L45} or \eqref{eqn:L24}.
Finding the global minimum is practically impossible, due to the large number of parameters (which is eleven). 
We thus take a statistical approach and proceed as follows.

\begin{description}
  \item[Sampling]We first generate a set of initial values for the parameters $x_i$, $i=0, 1, \cdots, 10$ by randomly sampling values from the uniform distribution for $0\leq x_i<2\pi$.
  \item[Optimisation]Then using a optimisation scheme of machine learning the loss function is optimised (minimised), for $N_{\rm iter}$ number of iteration steps.
  \item[Statistics]Repeating the process of sampling and optimisation, $N_{\rm samp}$ samples of numerical minimisation are collected. We then observe the minimised loss function values and the optimised configurations of the parameters $x_i$.
\end{description}

We summarise the obtained results in the next section.

\section{\label{sec:results}Numerical results}

We randomly generated 1000 initial configurations of the 11 parameters, and for each of these 1000 samples we optimised the parameters using the loss functions of the 45-Higgs model and the 24-Higgs model, up to 1,000,000 iterations.
For numerical minimisation of the loss functions \eqref{eqn:L45} and \eqref{eqn:L24}, we used the standard optimisation schemes of machine learning\footnote{
The results presented here are generated by the vanilla Adam algorithm \cite{Kingma:2014vow} with hyperparameters $\alpha=0.001$, $\beta_1=0.9$, $\beta_2=0.999$ and $\epsilon=10^{-8}$. 
Smaller learning rate ($\alpha$) is found to give smaller values of optimised loss functions (values averaged over 999,901st-1,000,000th data points).
Similar results are obtained by the Gradient Descent algorithm.
}.
%

\begin{figure*}
\includegraphics[width=85mm]{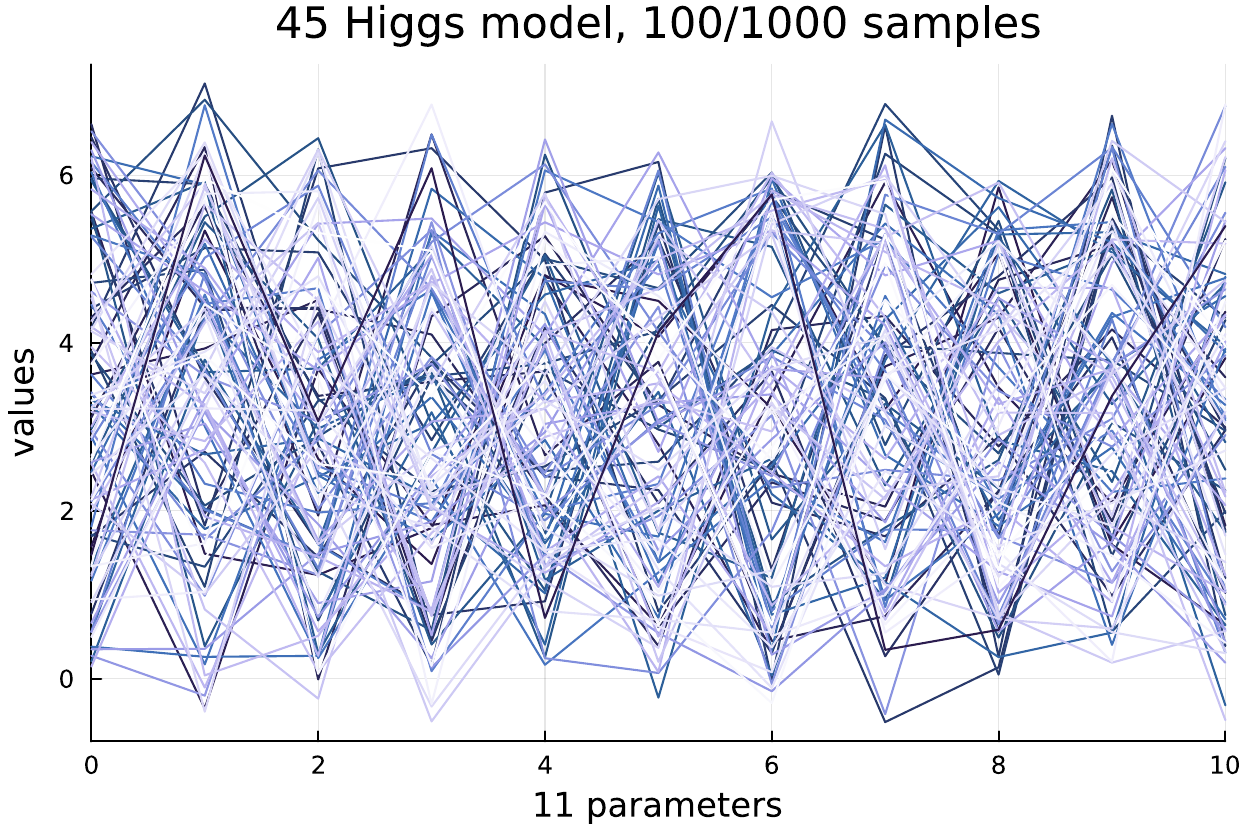}%
\includegraphics[width=85mm]{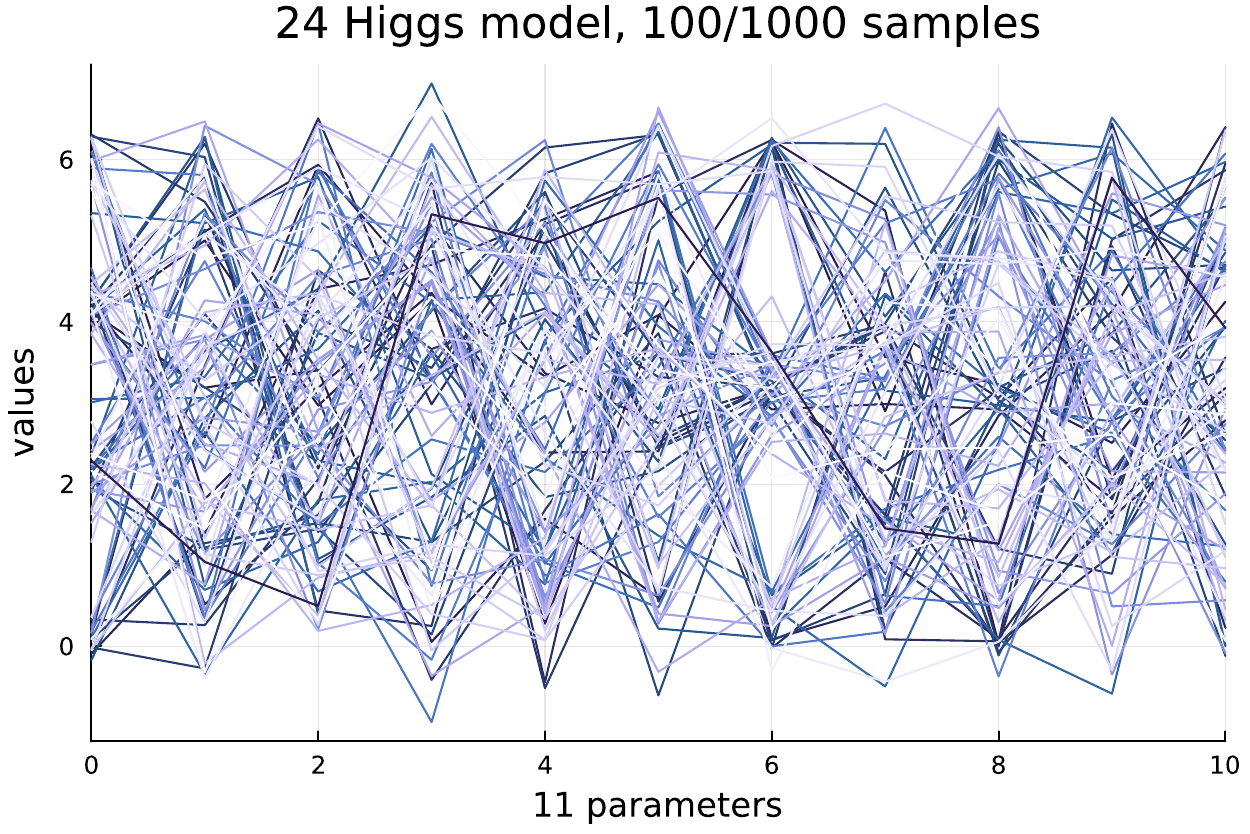}%
\caption
{\label{fig:ParamStat}
Configurations of the eleven parameters $x_i$, $i=0,\cdots 10$ after optimisation. 
The case of the 45-Higgs model is shown on the left and that of the 24-Higgs model is shown on the right.
Each panel displays 100 samples that give rise to the smallest 10\% of the optimised loss function values, out of 1000 samples.
Darker lines indicate smaller values of the optimised loss function.
}
\end{figure*}

\subsection{\label{sec:Loss}Loss function}%

Fig.~\ref{fig:LossEvo} shows a typical behaviour of the loss function in the process of optimisation, for the 45-Higgs model (blue) and the 24-Higgs model (red). 
In both models, the loss function reaches a plateau before $N_{\rm iter}\simeq 500,000$ iteration steps and the optimisation is seen to be completed; the decrease of the loss function afterwards is seen to be minor.

Typical minimised values of the loss functions for the 45-Higgs model and the 24-Higgs model are found to be different.
Fig.~\ref{fig:histogram} shows the distribution of the 1000 samples of optimised loss function values, for the two models with randomly chosen common initial configurations.
We evaluate these values by averaging over the last 100 (i.e. 999,901st to 1,000,000th) steps, since the loss functions exhibit spiky behaviour (which is common for this numerical algorithm).
The loss function of the 24-Higgs model tends to be optimised to smaller values than that of the 45-Higgs model. 
This indicates that statistically the 24-Higgs model can reach closer to the minimal $SU(5)$ model than the 45-Higgs model can. 
That is, in the parlance of \cite{Matchev:2024ash}, the 24-Higgs model is more beautiful than the 45-Higgs model.

\subsection{\label{sec:ParamOptim}Optimisation of parameters}%

In the process of optimisation, the parameters $x_i$ are adjusted to move so that the loss function takes smaller values.
A sample of the evolution of these parameters is shown in Fig.~\ref{fig:ParamEvo}, for the 45-Higgs model (the left penel) and the 24-Higgs model (the right panel).
Here, a common initial parameter set is chosen for the two models.
The parameter configurations evolve differently and settle to different optimised configurations. 
In both panels of Fig.~\ref{fig:ParamEvo}, the parameter set for the initial (randomly assigned) parameter values are shown in the darkest blue, and the (almost) optimised values at the $N_{\rm iter} = 1,000$th iteration step are shown in the brightest yellow, with the intermediate steps (10 iteration intervals) shown in the gradient colours.
After $N_{\rm iter}\simeq 1,000$, no appreciable change of parameters is seen in this sample.
The excursions of the 11 parameters are seen to be relatively small.
The optimisation algorithm does not explore the whole parameter space, but searches only the vicinity of the initial configuration and finds a local minimum nearby.
This indicates that the loss function has many local minima distributed across the 11 dimensional parameter space.

Finally, Fig.~\ref{fig:ParamStat} shows the optimised configurations of parameters, for the 45-Higgs model (left) and the 24-Higgs model (right).
We have chosen 100 samples that give the smallest 10\% of the optimised loss function out of 1000 collected samples. 
Darker lines correspond to smaller values of the optimised loss function.
It is seen that, in particular for $x_5$-$x_8$ corresponding to the CKM-like nondiagonal part of our parametrisation \eqref{eqn:CKMlike}, dark lines are seen to congregate and there appear some blank regions.
The blank regions are apparently disfavoured in the sense that the loss function is not optimised to a very small value, suggesting that these parameters are not beautiful. 
The contrast is more evident in the 24-Higgs model than in the 45-Higgs model.


\section{\label{sec:final}Final remarks}

The minimal $SU(5)$ GUT model 
fits all known particles into simple representations of the $SU(5)$ gauge group.
It certainly is a beautiful theory, but does not harmonise with experimentally supported, truth.
By extending the theory, either with the $\bm{\overline{45}}$ representation Higgs or including the higher dimensional operator involving the $\bm{24}$ representation Higgs, the theory can be reconciled with the truth.
These extensions of the $SU(5)$ model introduce many degrees of freedom and the model suffers from a curse of dimensionality; a comprehensive parameter search is practically impossible.
In this paper we analysed these two extensions of the $SU(5)$ GUT model, making use of techniques from machine learning. 
Instead of performing an exhaustive parameter search, we introduced a criterion of {\em goodness} of parameters, which is encoded in the form of the loss function, and then explored the parameter space through sampling and optimisation.
The results indicate that overall, the 24-Higgs model gives smaller values of loss function after optimisation, compared to the 45-Higgs model.
This indicates that the 24-Higgs model can reach closer to the original $SU(5)$ model

Let us conclude with some comments on possible directions of future research.
While a common attitude in machine learning is data-driven, that is, to avoid human intervention 
on data processing wherever possible, we had to make a choice of the loss function in which our prejudice on good parameters is reflected.
We chose the specific form of the loss function, which we think is reasonable and also is suitable for optimisation by the back propagation algorithm of machine learning.
Our loss function of the determinant form (\eqref{eqn:L45} and \eqref{eqn:L24}) treats the contributions from the three families equally. 
Concerning the $SU(5)$ GUT mass relation, it is known that the disagreement in the third generation (the bottom mass and the tau mass) is not as intense as the first and the second generations (the down mass-electron mass and the strange mass-muon mass).
If one wants to discuss family-dependent features of mass relations, the loss function needs to be suitably modified so that weights on generations are included.

The second point concerns the choice of parameters. 
We fixed the mass matrices $M_e$ and $M_u$ using the experimental central values and parameterised $M_d$ with 11 parameters. 
This is an economical choice that fully covers the flavour sector while respecting current observational constraints. 
As we examine Fig.~\ref{fig:ParamStat}, the appearance of blank regions may suggest some parameters are disfavoured.
There might be a better way to parameterise the flavour sector that makes such features more evident.

Finally, it would be interesting to investigate whether the {\em goodness} of the flavour sector parameters, as discussed in terms of the loss function, has any implications on the phenomenology of grand unification, which includes aspects, such as neutrino physics, baryogenesis, proton lifetime \cite{Dorsner:2024seb} and successful realisation of cosmic inflation \cite{Guth:1980zm,Arai:2011nq,Kawai:2015ryj}.

\begin{acknowledgments}
S.K. acknowledges warm hospitality of the Department of Physics and Astronomy, University of Alabama and Helsinki Institute of Physics and the Department of Physics, University of Helsinki. 
This work was supported in part by 
the National Research Foundation of Korea Grant-in-Aid for Scientific Research Grant No. NRF-2022R1F1A1076172 (S.K.) and
by the United States Department of Energy Grant Nos. DE-SC0012447 and DE-SC0023713 (N.O.).
\end{acknowledgments}




%

\end{document}